\documentclass[sigconf]{acmart}

\AtBeginDocument{%
  \providecommand\BibTeX{{%
    \normalfont B\kern-0.5em{\scshape i\kern-0.25em b}\kern-0.8em\TeX}}}
    
\copyrightyear{2021}
\acmYear{2021}
\setcopyright{acmcopyright}
\acmConference[SIGIR '21] {Proceedings of the 44th International ACM SIGIR Conference on Research and Development in Information Retrieval}{July 11--15, 2021}{Virtual Event, Canada.}
\acmBooktitle{Proceedings of the 44th International ACM SIGIR Conference on Research and Development in Information Retrieval (SIGIR '21), July 11--15, 2021, Virtual Event, Canada}
\acmPrice{15.00}
\acmISBN{978-1-4503-8037-9/21/07}
\acmDOI{10.1145/3404835.3464927}

\usepackage{subcaption}
\usepackage{graphicx}
\usepackage{algorithm}  
\usepackage{algorithmic}
\usepackage[skip=3pt]{caption}
\usepackage[compact]{titlesec}

\newcommand{\eg}{\textit{e.g.}}

\renewcommand{\G}{\mathcal{G}}
\newcommand{\V}{\mathcal{V}}
\newcommand{\E}{\mathcal{E}}
\newcommand{\N}{\mathcal{N}}
\newcommand{\Q}{\mathcal{Q}}
\newcommand{\I}{\mathcal{I}}
\newcommand{\h}{\mathbf{h}}
\newcommand{\w}{\mathbf{w}}
\newcommand{\z}{\mathbf{z}}

\newcommand{\header}[1]{{\flushleft \textbf{#1}}}

\newcommand\blfootnote[1]{%
  \begingroup
  \renewcommand\thefootnote{}\footnote{#1}%
  \addtocounter{footnote}{-1}%
  \endgroup
}

\newcommand{\sgcn}{SearchGCN}

\begin{document}
\fancyhead{}

\title{SearchGCN: Powering Embedding Retrieval by Graph Convolution Networks for E-Commerce Search}

\author{Xinlin Xia$\,^{1\dagger}$, Shang Wang$\,^{1\dagger}$, Han Zhang$\,^{1}$, Songlin Wang$\,^{1}$, Sulong Xu$\,^{1}$, Yun Xiao$\,^{2}$, Bo Long$\,^{1,2}$, Wen-Yun Yang$\,^{2*}$}
\affiliation{%
  \institution{$^{1}\,$JD.com, Beijing, China; $^{2}\,$JD.com Silicon Valley Research Center, Mountain View, CA, United States}
  \{xiaxinlin3, wangshang1, zhanghan33, wangsonglin3, xusulong, xiaoyun1, bo.long, wenyun.yang\}@jd.com
}

\renewcommand{\shortauthors}{Xinlin Xia, et al.}

\ccsdesc[500]{Computing methodologies~Neural networks}
\ccsdesc[500]{Information systems~Information retrieval}

\begin{abstract}
 Graph convolution networks (GCN), which recently becomes new state-of-the-art method for graph node classification, recommendation and other applications, has not been successfully applied to industrial-scale search engine yet. In this proposal, we introduce our approach, namely {\sgcn}, for embedding-based candidate retrieval in one of the largest e-commerce search engine in the world. Empirical studies demonstrate that {\sgcn} learns better embedding representations than existing methods, especially for long tail queries and items. Thus, {\sgcn} has been deployed into JD.com's search production since July 2020.
\blfootnote{$^\dagger\,$ Both authors contribute equally}
\blfootnote{$^*\,$ Corresponding author at wenyun.yang@jd.com}
\end{abstract}

\keywords{Search; Neural networks; Graph convolution networks; Representation learning}

\maketitle

\section{Introduction}

E-commerce search, which serves as an essential part of online shopping platforms, typically consists of three steps: query processing, candidate retrieval and ranking. In this paper, we focus solely on the candidate retrieval step which aims to retrieve thousands of candidates from billions of items.

Despite the recent advance of embedding based retrieval methods~\cite{zhang2020towards} and their successful applications in candidate retrieval stage, in practice we still notice a few drawbacks: 1) representations of low-frequency items are not well learned due to their rare occurrences; 2) Query embeddings are computed purely from a handful of query tokens, which sometimes can not capture the query semantic meanings well; 3) The training efficiency is low, mainly because that each training example consists of only one pair of query and clicked item.

\begin{figure}[!t]
    \centering
    \hspace*{4mm}
    \includegraphics[width=0.45\textwidth]{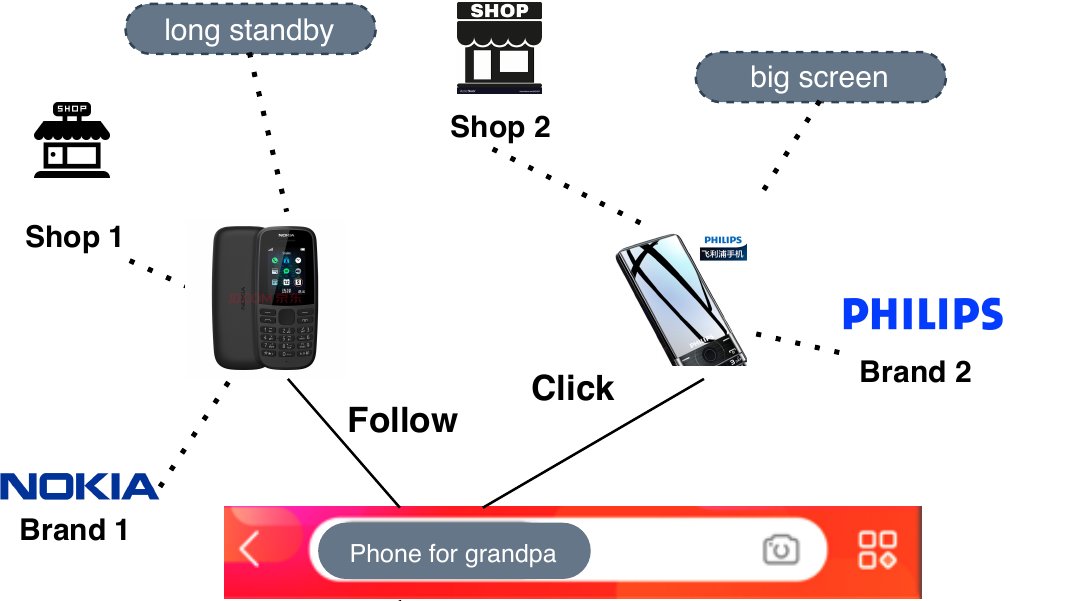}
    \caption{Illustration of e-commerce search graph.}
    \label{fig:ecomm_graph}
    \vspace{-6mm}
\end{figure}

In this paper, we explore Graph Convolution Networks (GCN) to enrich entity representations by aggregating related neighbor's information. In our case of e-commerce search, it is very likely that we can learn a better embedding representation for a user query, if we can aggregate information from the most clicked or bought items for the query to alleviate the shortage of query tokens. Likewise, we can learn a better item embedding representation, if we can aggregate information from the most frequent queries that lead to the item clicks or orders. Figure~\ref{fig:ecomm_graph} shows a quick glance of typical e-commerce graph  which illustrates how a query aggregates information from related items. Moreover, we explore how to 
improve the training efficiency by neighbor sampling.

\section{Method}
\label{sec:method}


Given a graph $\G = (\V, \E)$ where each $v\in \V$ stands for an entity node, and each $e \in \E$ stands for an edge between two entities, let's denote the set of all the neighbors for a node $v$ as $\N_v$, the set of queries as $\Q$, and the set of items as $\I$, where $\Q \subseteq \V$ and $\I \subseteq \V$.

\header{Neighbor Aggregation.}
The core of GCN is to compute the $l$-th layer's representation of a target node by all its neighbors and itself in $(l-1)$-th layer. Formally, we can define the aggregation operation as follows
\begin{equation*}
\h_{v}^{l} = \text{AGGREGATE}(\h_v^{l-1}, \h_{u}^{l-1}, \forall u \in \mathcal{N}_{v} ),
\end{equation*}
where $\h_v^l\in \mathbb{R}^{d}$ denotes the $d$-dimensional embedding representation of a node $v$ at $l$-th layer, the AGGREGATE function is a pooling function that takes the target node and all its neighbor nodes as inputs. 

In \sgcn, we use the following neighbor aggregation function that can be explained as a weighted sum of all its neighbors,
\begin{equation}
\h_{v}^{l} = \sum_{u \in \mathcal{N}_{v}} \alpha^{l-1}_{vu} \h_{u}^{l-1},
\label{eq:attention_aggregation}
\end{equation}
\begin{equation}
\alpha^{l-1}_{vu} = \frac{\text{exp}\left(\text{LeakyReLU}\left({\w^{l-1}}^\top [\h^{l-1}_{v}||\h^{l-1}_{u}]\right) \right)} 
{\sum_{u\in\mathcal{N}_{v}} \text{exp} \left( \text{LeakyReLU}\left({\w^{l-1}}^\top [\h^{l-1}_{v}||\h^{l-1}_{u}]\right) \right)},
\label{eq:attention_weight}
\end{equation}
where $.^\top$ denotes the transpose operation, $||$ denotes the concatenation operation and $\mathbf{w}^{l-1} \in \mathbb{R}^{2d}$ is model parameter to learn. 

\header{Layer Combination.}
In our {\sgcn}, to refine the embedding of query and item node, we sum the neighbor embeddings obtained by neighbor aggregation layer-by-layer to form the final embedding:
\begin{equation}
\z_{v} = \sum_{l=0}^{L} \h_{v}^{l}. \nonumber
\end{equation}

\header{Neighbor Sampling.}
Since our e-commerce graph has hundreds of millions of nodes and billions of edges, which is too large to be fully loaded into computer memory, and the e-commerce graph appears to be significant imbalance,  which makes the neighbor aggregation expensive. Therefore, to be practical in an industrial system, we need to develop a neighbor sampling approach that is efficient in both memory and computation.

First, we prune the neighborhood size for each node to at most $50$, by the weights of the edges between each neighbor and the target node, to allow us to load all edges into memory during the training. Second, we introduce randomness to neighbor sampling, by sampling some fixed number (\eg, $10$) of neighbors according to edge weights, to reduce the computational cost to aggregate neighbors and increase model generalization ability.

\section{Experiments}
\label{sec:experiment}
We conduct experiments on a data set of $60$ days' user click logs and we can assume that if an item is clicked for a query, the item is relevant to the query.

\header{Offline Evaluation.}
We compare {\sgcn} with a few baseline methods and report the evaluation results in Table~\ref{tab:offline_comparison}. \textit{SearchGCN-mean} adopts mean aggregator that gives equal weight to all neighbors. \textit{SearchGCN-attention} adopts attention mechanism in aggregation function as shown in Eqs. (\ref{eq:attention_aggregation}) and (\ref{eq:attention_weight}). \textit{SearchGCN-mask} adopts neighbor masking on both query and item side to mask the target query and item.

\begin{table}[tb]
    \centering
    \caption{Comparison between different methods.}
    \label{tab:offline_comparison}
    \begin{tabular}{c|cccc}
    \hline
        & Top-$1$   & Top-$10$   & AUC   & Sec./Step \\
    \hline
    DPSR~\cite{zhang2020towards} & $86.23\%$ & $99.59\%$ & $0.737$ & $0.276$ \\
    GraphSAGE~\cite{hamilton2017inductive} & $79.29\%$ & $95.97\%$ & $\mathbf{0.783}$ & $1.03$ \\
    PinSAGE~\cite{ying2018graph} & $68.59\%$ & $93.18\%$ & $0.736$ & $1.07$ \\
    \hline
    {\sgcn}-mean & $86.76\%$ & $99.43\%$ & $0.764$ & $1.04$ \\
    {\sgcn}-attention & $86.57\%$ & $99.27\%$ & $0.751$ & $1.05$ \\
    {\sgcn}-mask & $\mathbf{87.72\%}$ & $\mathbf{99.67\%}$ & $0.768$ & $0.98$ \\
    \hline
    \end{tabular}
\end{table}

\header{Online A/B Tests.} We conducted live experiments over $10\%$ of the entire site traffic during a period of one week using a standard A/B testing configuration. To protect confidential business information, only relative improvements are reported. Table~\ref{tab:abtest} shows that the proposed {\sgcn} retrieval improves the production system for core business metrics in e-commerce search, including user conversation rate (UCVR), revenue per mille (RPM) and gross merchandise value (GMV). As a result, the proposed {\sgcn} has been deployed into JD.com's search production since July 2020.

\begin{table}[t]
\centering
    \caption{Improvements of {\sgcn} in online A/B test.}
    \label{tab:abtest}
    
    \begin{tabular}{c|r r r r}
    \hline
            & UCVR  & RPM & GMV  \\
    \hline
    SearchGCN  & $+0.20\%$ & $+1.13\%$ & $+1.27\%$  \\
    \hline 
    \end{tabular}
\end{table}

\header{Improvement on Torso and Tail Query.}
We evaluate {\sgcn}'s performance on different types of queries. We split our evaluation data set into three categories, head, torso and tail queries, according to the queries' frequency. 
As shown in Table~\ref{tab:tail_query}, we can see that the improvements on torso and tail queries are more significant than head queries. This observation strongly suggests that the GCN's neighbor aggregation could benefit more for the torso and tail queries, potentially because they can learn from their neighbor's representations. Similar trend also holds for long tail items, but we have to skip due to space limit.

\begin{table}[t]
    \centering
    \caption{Comparison between different methods on head, torso, and tail queries. The metric is top-1 score.}
    \label{tab:tail_query}
    \begin{tabular}{c|ccc}
    \hline
      & Head & Torso  & Tail \\
    \hline
    DPSR & $84.23\%$ & $92.61\%$ & $92.64\%$ \\
    SearchGCN & $85.73\%$ & $94.68\%$ & $94.77\%$ \\
    \hline
    Difference & $1.50\%$ & $2.07\%$ & $2.13\%$ \\
    Error reduction rate & $9.51\%$ & $28.01\%$ & $28.94\%$ \\
    \hline
    \end{tabular}
    \vspace{-4mm}
\end{table}

\bibliographystyle{ACM-Reference-Format}
\balance

\bibliography{references}

\appendix

\section{Presenter Profiles}
\header{Xinlin Xia} is a researcher in the Department of Search and Recommendation at JD.com Beijing. He received his master degree in Computer Science from Sichuan University, China. His research focuses on information retrieval.

\header{Han Zhang} is a researcher in the Department of Search and Recommendation at JD.com Beijing. She graduated from Natural Language Processing Lab, Department of Computer Science and Technology, Tsinghua University. Her research focuses on information retrieval.

\header{Wen-Yun Yang} is a principal scientist at JD.com and previously a research scientist at Facebook News Feed team. He had about 8 years experience of tech leading in search and recommendation backend systems, candidate retrieval, ranking and modeling in general. He received his Ph.D. degree in Computer Science from University of California, Los Angeles.

\end{document}